\documentclass[a4]{article}

\usepackage{lineno}
\usepackage{amssymb}
\usepackage{amsmath}
\usepackage{color}
\usepackage{colortbl}
\usepackage{verbatim}
\usepackage[round]{natbib}
\usepackage{multirow}
\usepackage{ulem}
\usepackage{setspace}
\usepackage{rotating}
\usepackage[blocks]{authblk}
\usepackage{url}
\usepackage{graphicx}
 \usepackage{fullpage}
\usepackage{inputenc}
\usepackage{threeparttable}

\usepackage[a4paper]{geometry}
\linenumbers*[1]

\begin{document}


\thispagestyle{empty}



\title{\textbf{Anthropogenic Renourishment Feedback on Shorebirds: a Multispecies Bayesian Perspective}}

\author{\textbf{M. Convertino$^{\S^*}$, J.F. Donoghue $^{\sharp}$, M.L. Chu-Agor$^{\S}$, G. A. Kiker$^{\S}$, R. Mu\~{n}oz-Carpena$^{\S}$, R.A. Fischer$^{\ddag}$, I. Linkov$^{*\dag}$}}

\vspace{1cm}

\affil{$^{\S}$ Department of Agricultural and Biological Engineering-IFAS, University of Florida, Gainesville, FL, USA}

\affil{$^{*}$ USACE Engineer Research and Development Center (ERDC), Risk and Decision Science Area, Concord, MA, USA} 

\affil{$^{\sharp}$ Department of Earth, Ocean and Atmospheric Science, Florida State University, Tallahasse, FL, USA}

\affil{$^{\ddag}$ USACE Engineer Research and Development Center (ERDC), Environmental Laboratory, Vicksburg, MS, USA}

\affil{$^{\dag}$ EPP Dept. Carnegie Mellon University, Pittsburgh, PA, USA}

\maketitle
\thispagestyle{empty}

{\it Corresponding author:} M. Convertino, Department of Agricultural and Biological Engineering-IFAS, Frazier Rogers Hall, Museum Road, PO box 110570, 32611-0570, (email: mconvertino@ufl.edu; phone: +1 352-392-1864 ext. 218; fax: +1 352-392-4092).\\


\doublespacing

\newpage
\begin{abstract} \label{abstract}

In this paper the realized niche of the Snowy Plover (\textit{Charadrius alexandrinus}), a primarily resident Florida shorebird, is described as a function of the scenopoetic and bionomic variables at the nest-, landscape-, and regional-scale. We identified some possible geomorphological controls that influence nest-site selection and survival using data collected along the Florida Gulf coast. In particular we focused on the effects of beach replenishment interventions on the Snowy Plover (SP), and on the migratory Piping Plover (PP) (\textit{Charadrius melodus}) and Red Knot (RK) (\textit{Calidris canutus}). Additionally we investigated the potential differences between the SP breeding and wintering distributions using only regional-scale physiognomic variables and the recorded occurrences. To quantify the relationship between past renourishment projects and shorebird species we used a Monte Carlo procedure to sample from the posterior distribution of the binomial probabilities that a region is not a nesting or a wintering ground conditional on the occurrence of a beach replenishment intervention in the same and the previous year. The results indicate that it was 2.3, 3.1, and 0.8 times more likely that a region was not a wintering ground following a year with a renourishment intervention for the SP, PP and RK respectively. For the SP it was 2.5. times more likely that a region was not a breeding ground after a renourishment event. Through a maximum entropy principle model we observed small differences in the habitat use of the SP during the breeding and the wintering season. However the habitats where RK was observed appeared quite different. While ecological niche models at the macro-scale are useful for determining habitat suitability ranges, the characterization of the species' local niche is fundamentally important for adopting concrete multispecies management scenarios. Maintaining and creating optimal suitable habitats for SP characterized by sparse low vegetation in the foredunes areas, and uneven/low-slope beach surfaces, is the proposed conservation scenario to convert anthropic beach restorations and SP populations into a positive feedback without impacting other threatened shorebird species.

\end{abstract}





\section{Introduction} \label{intro}

The increasing availability of spatio-temporal data on species presence, along with the availability of remotely sensed data and GIS techniques, has greatly enhanced in the last decade the study of the distribution of thousands of species \citep{elith06,soberon07}. However, the individuation of the species' range is performed in a Grinnellian way, considering mostly scenopoetic variables that are suitable to describe the fundamental niche \citep{soberon07}. Studies on the distribution of species require the consideration of biotic variables (Eltonian perspective) in order to truly characterize the realized and the fundamental niches \citep{colwell09}. If the habitat exhibits conditions that lie entirely within a species' niche, a population persists without immigration from the external world, whereas if conditions lie outside the niche, the species faces possible extinction. Analysis of species' niches are essential to understand controls on species' geographical range limits and how these limits might shift in response to climatic changes \citep{holtufl09,tingley09,zimmermann09}. The Hutchinson' s duality consists in considering simultaneously exogenous variables that describe the biotope in which the species live, and the biotic variables that characterize the interactions of the species with other living and non-living controls \citep{colwell09}. Recently, the emerging fields of phylogeography and landscape ecology \citep{knowles09,wangphilo10} have significantly improved species distribution modeling and to detect differences among species including data at the cell-level (e.g. DNA sequences) \citep{funk07,rissler07,kupper09,kearney09,usgsmolpp09}. \\

\begin{figure}
\begin{center}
\advance\leftskip-1.5cm
\includegraphics[width=16cm]{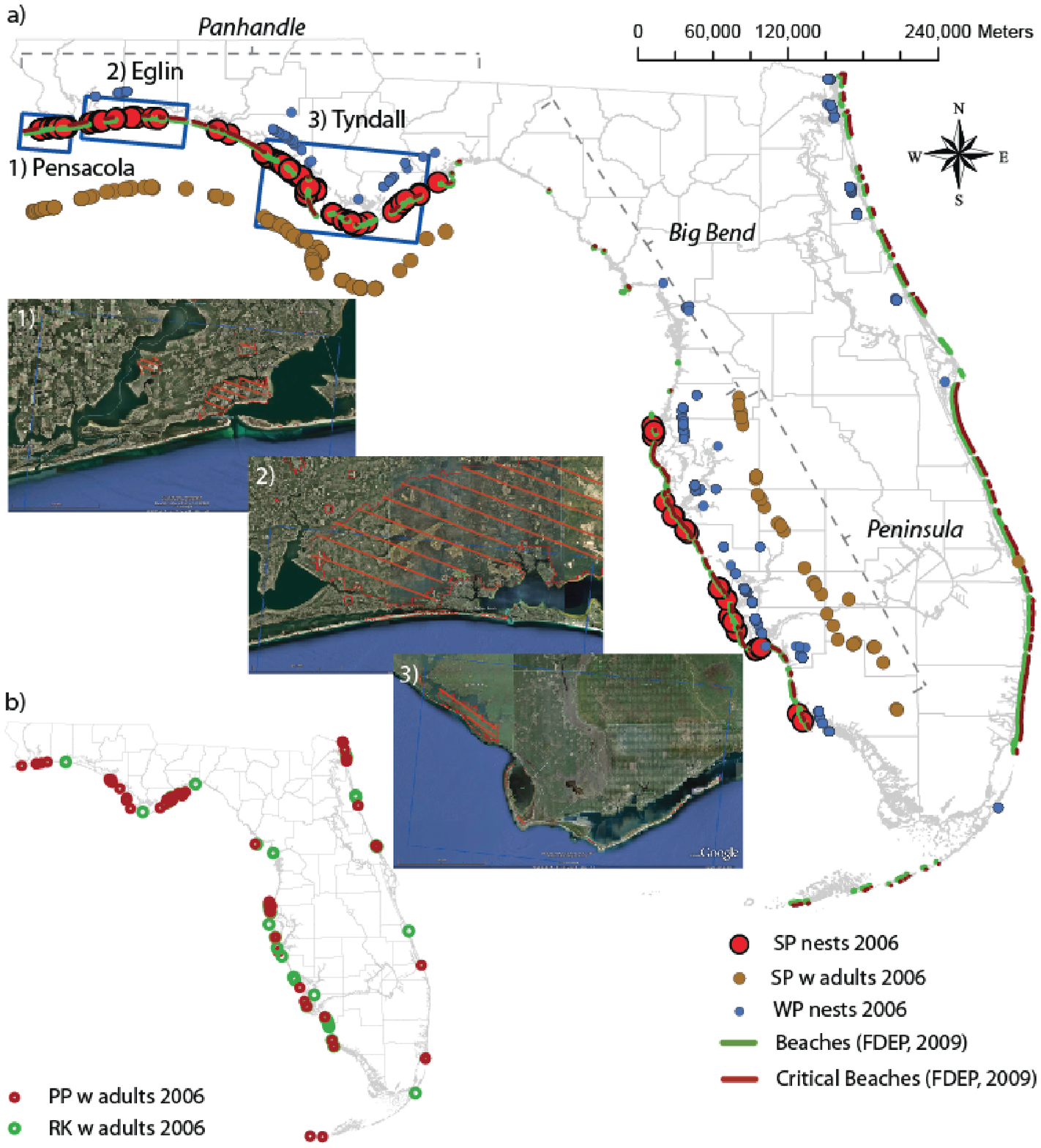}
\caption[h]{} \label{fig1}
\end{center}
\end{figure}

The selection of breeding and wintering habitat by shorebirds and their consequent survival may be influenced by a combination of factors, including human recreational activities, predator activity, prey availability, and the habitat substrate \citep{hoover98,newton98,jones01,colwell07a}. Nest-site selection and nest-survival patterns reveal in general an influence by a combination of the aforementioned environmental and biological factors working in concert in addition to physical features surrounding the nest-site. Few avian habitat studies have been able to compare multiple ecological hypotheses of species distribution and multispecies nest-site selection decisions \citep{jones01} to aide management policies. While habitat selection is often assumed to be adaptive, evidence for adaptive habitat selection in birds has been mixed \citep{clark99,jones01}. The consideration of multiple predictors collectively for detecting species distribution, is useful for habitat management, and it benefits the conservation of rare and declining species.
Shorebirds reproductive success is correlated with the stability and quality of the nesting environment. In particular, here we study the effects of beach replenishment (or renourishment) on the Snowy Plover (\textit{Charadrius alexandrinus}), a state-threatened shorebird in Florida (Figure \ref{fig1}). SP females in general show fidelity to nesting beaches, making artificial beach nourishment practices and the subsequent physical and biological changes to habitat directly relevant to their recovery. The reduction in reproductive output is in general primarily a consequence of decreased nesting success. The result of reduced nesting success is more precisely described as reduced juvenile recruitment rather than reduced number of nests, since nesting attempts still occur but are not successful. However late renourishments in the wintering season can be a source of disturbance for the Snowy Plover. In comparison to the SP, two other shorebird species have been considered. The Piping Plover, \textit{Charadrius melodus} (federally designated as threatened), and the Red Knot, \textit{Calidris canutus} (threatened in New Jersey, and a candidate for Endangered Species Act protection), are migratory shorebirds whose wintering/stopover time in Florida is on average 3 months and 3 weeks respectively \citep{harri01,smith04ok} (Figure \ref{fig1}). The Red Knot subspecies \textit{Calidris canutus rufa} is the only endangered species among the species considered, under the federal Endangered Species Act. It has been established that the \textit{Calidris canutus rufa} uses some Florida beaches as stopover areas during its migratory route to South America. The Wilson's Plover (WP) is an other resident shorebird reputed to be the main competitor of the SP, however it is not considered in this study due to its least-concern status in Florida. \\

Due to habitat loss, many threatened, endangered, and at-risk species (TER-S) are decreasing in abundance. The a-priori evaluation of the effect of beach restoration activity on species is fundamentally important to understanding the effectiveness of the intervention and to optimizing strategies. Beach renourishment is mainly carried out to preserve existing structures and to increase the beach area \citep{smith09}. This potentially translates into new income from tourism and beach activities. In Florida, an average of $\$$ 90 million is spent annually on beach renourishment \citep{psds10}. Over the past two decades, more than 50 large renourishment projects have been undertaken in the state, with a typical project averaging approximately 4.5 km in length.  More than 242 km of Gulf and Atlantic coast beaches have been impacted by renourishment sand during that time \citep{wang05}. The five Gulf states account for more than forty percent of all renourishment activity in the United States \citep{psds10}.  Florida alone accounts for thirty percent \citep{finkl96}. Beach renourishment efforts are not without significant social or legal impacts. In June 2010 the Supreme Court handed down a unanimous verdict that effectively allows the Florida state government to resume beach renourishment projects without paying for property that homeowners claims have been ``taken''. In 2003 a group of NW Florida coastal homeowners protested against a replenishment, claiming lowered property values due to the increase in beach width (on average 50 m) and subsequent greater public access to the beach. The US Supreme Court rejected the appeal, declaring the renourishment as a necessary intervention for preserving the coastal ecological communities and human structures especially in light of the increase in sea-level rise and of extreme meteorological events due to climate change \citep{supreme10}. As a societal issue, beach renourishment is one of the most expensive interventions in civil and environmental engineering, considering also the environmental variables that are often unpredictable and the factors affecting the coastline due to climate change and extreme climatological events \citep{smith09,gopalakrishnan10,landry2010}. For instance unpredictable variations in ocean energy impact due to littoral currents, or strong hurricanes, can rapidly destroy the renourished areas. Renourishment projects are rarely designed to incorporate the life-history needs of shoreline-dependent species \citep{nourish10}. The primary considerations in planning a renourishment are sand source, and compatibility of the borrowed sand with the native beach, including grain size, composition and color. Disturbances associated with beach replenishment, such as dredging and sand-pipeline movement, represent an additional and potentially significant barrier to breeding and nesting for both shorebirds and waterbirds. For example, investigators have recommended avoiding beach management practices that disturb beach microhabitats (e.g., ephemeral pools and bay tidal flats) important for Snowy Plover and Piping Plover chick survival \citep{elias00,grippo07}. Replenishment entails substantial changes in beach morphology that potentially cause changes in local movement patterns, resting behavior, or habitat use of shorebird species during the tidal cycle. Similarly, the potential disturbance to benthic macroinvertebrate assemblages could alter feeding behavior in bird species whose diet relies on benthic organisms \citep{bishop05,dugan06,peterson06marine}. Beach renourishment has been found to alter shorebird distributions more than seabird distributions \citep{grippo07}. However not many studies exist on the topic and the effect of renourishments seems to be very strongly species-dependent \citep{grippo07}. \citet{lott09} analyzed the effect of sand replacement projects on SP and PP along the Florida beaches reporting a qualitative negative correlation, without indication of the causes. \citet{jackson10} found that beach renourishment programs in estuaries can enhance shore protection, but can decrease habitat suitability by changing the beach shape creating higher berms and wider backshores than would occur under natural conditions. However \citet{jackson10} did not focus on any species in particular. The controversy about renourishment vs. species-abundance and other species patterns, has also involved other shoreline dependent taxa. For example, \citep{brock07} studied the influence of beach replenishment on sea-turtles finding a clear decline in the nesting success and abundance in the season after the anthropic restoration. \citet{mennpap02,menn02b,asmfc06,erdcren06,delahuz08} are in agreement that most of the actual beach renourishment projects worldwide disturb the food-web structure of the coastal habitat ecosystem impacting all the species occupying the affected niche. \citet{mennpap02,menn02b}, and \citet{delahuz08} noticed the evident linkage between the geomorphodynamic structure of the beach and the quality of the habitat in sustaining species (from microorganisms to birds). Dredging for beach renourishments was found also to impact coral reef communities \citep{jaap00}. The three factors of beach intervention costs, biodiversity protection, and potential income, make the a-priori adaptive management \citep{thom00}, risk assessment, uncertainty and decision analysis of renourishment fundamentally important \citep{nordstrom05}. \\

We described in relation to beach renourishment, the role of prey availability, predator activity, human activity, and physical features on habitat use and possible survival of the Snowy Plover (\textit{Charadrius alexandrines}) in Florida. For Florida, many technical reports about shoreline-dependent birds, in particular Snowy Plover, have been produced \citep{gore89,lamonte02,himes06,fwcca07,burney09,ippc09,lott09,pruner10}. However there is still a lack of quantitative studies addressing the ecological effects of renourishment on shoreline-dependent birds. Although the relationship between the use of coastal habitat and shorebirds has been assessed in much detail by many previous studies, e.g. \citet{taft06,hood07,hood07b,sirami08,tian08,gan09}, as well as for riverine ecosystems in proximity of the coast \citep{colwell05}, the literature about habitat suitability modeling for shorebirds is not extensive. Recently, the Snowy Plover has been studied intensively as part of a joint effort of the US Department of Defense, the Environmental Protection Agency, and the Department of Energy, for its conservation as a function of climate change and military activity \citep{chuagor10,convertino10time,convertino10hur,convertinoNATO,convertino10scalesbirds, convertino10uncert}. \citet{convertino10scalesbirds} provided evidence of the fundamental niche of SP through a maximum entropy approach, that is composed by the estuarine and ocean beaches constituted of alkaline medium/fine white sand and silt. \citet{convertino10uncert} described the source of uncertainty in data and species distribution models for the particular case of the SP in Florida. \citet{convertino10hur} found an interesting interannual positive feedback between the tropical cyclones in the year prior to a breeding season and the SP abundance and range. \citet{chuagor10} modeled the habitat evolution of Santa Rosa Island along the Florida Panhandle, and \citet{convertino10time} identified a decline in the power-law distribution of the habitat patch-size (i.e. the probability of finding suitable breeding/wintering habitat patches larger than a given size) for the SP, PP, and RK, through coupled modeling of the land-cover and of the habitat suitability as a function of the IPCC A1B sea-level rise scenario rescaled to 2 m of sea-lever rise. Recently \citet{Seavey2010} studied the threat of the Piping Plover as a function of the sea-level rise due to climate change in their New York barrier islands habitat. These studies confirm the importance of identifying effective interventions, such as renourishments, that support the wildlife needs in the face of climate change. The purposes of this paper are to:

\begin{itemize}

\item Provide a comprehensive overview of the biology of the Snowy Plover in Florida in relation to other SP populations in the USA, and with other shorebirds in Florida, that is potentially useful in metapopulation modeling; 

\item Describe probabilistically the effect of past replenishments on the SP breeding/wintering range and abundance, together with the determination of the local-scale geomorphological features of the habitat that increase the probability of site-selection and survival. A comparison is performed with other TER-s species in Florida (PP and RK); 

\item Assess some management scenarios (specifically ``ecologically sustainable'' costal restorations) as a function of the environmental cues of the SP to reduce the risk of nest-failure. This lays the foundation for subsequent multi-criteria decision analysis for the conservation of the SP. 

\end{itemize}

\section{Materials and Methods} \label{data}

\subsection{Models}

We employed two models: (1) a maximum entropy approach model ({\sc MaxEnt}) \citep{phillips06,phillips08b,Elith10sum} to quantify the similarities in the habitat use of the Snowy Plover during the breeding and wintering seasons, and to evaluate the habitat preferences of other TER shorebird species; and, (2), a Bayesian approach model to evaluate the probability of coarse-scale site-selection for different species after replenishment events, sampling from the posterior probability and based on historical data. A detailed description of the biological data used in the models is contained in the Supplementary Material of the manuscript.\\

The maximum entropy model adopted ({\sc MaxEnt}) is fully described in \citep{phillips06} and \citep{phillips08b}. {\sc MaxEnt} has already been applied in modeling the habitat suitability of the SP during the breeding season by \citet{convertino10scalesbirds} and \citet{convertino10uncert}. The estuarine and ocean beaches composed of alkaline medium/fine white quartz sand and silt were found to be the most suitable for SP during the breeding season. Here we used the same approach, considering as explanatory variables at the regional scale (the entire Gulf coast of Florida) the land-cover and the geology map. The land-cover from NOAA  \citep{noaa93} has been translated into SLAMM land-cover classes. SLAMM (Sea Level Affecting Marshes Model) \citep{clough06} is the model we used to simulate the effect of sea-level rise on the coastal habitat \citep{chuagor10,convertino10time}. Tables 1 and 2 in \citet{convertino10scalesbirds} report in detail the land-cover and the geology classes. The fundamental niche for Piping Plovers and Red Knot is determined by {\sc MaxEnt} \citep{phillips06,phillips08b,Elith10sum} for the same geographical domain of the SP and with the same environmental variables (Figure \ref{fig4}). Occurrences for the SP, PP, and RK are for the 2006 wintering distribution. The habitat suitability maps are the average over 30 replicates performed for at least 10,000 random background points. \\

To quantify the apparent relationship between beach nourishment projects and SP, PP, and RK depicted in Figure \ref{fig5} we used a Monte Carlo procedure to sample from the posterior distribution of the binomial probabilities that a region is a nesting or wintering ground conditional on whether or not the same year or the previous year the region experienced a renourishment event.  Table \ref{table1} lists the number of nests by year in the breeding and in the wintering season, the number of renourishment events, and the average number of fledglings. For example, the data in the gray background in Table \ref{table1} are those considered for the Bayesian inference in the seasons 2005-2006 (the renourishment interventions in 2004 are considered when analyzing the 2005 breeding season). In the years 2008-2010 there were renourishment events in the three regions considered (Pensacola/Eglin, Tyndall, and Peninsula, see Figure \ref{fig1}). Data about the 2008-2010 renourishments are reported in the Supplementary Material.  The median number of nests per year for the SP, per region is 57. Here we define a nesting ground as an area having at least 10 nesting sites. A wintering ground is an area in which at least 2 adult individuals were observed. The threshold of 10 nesting sites is found to be a reasonable value that considers the breeding success and the minimum breeding area \citep{convertino10hur}. Because the solitary behavior of shorebirds \citep{convertino10scalesbirds} the occurrence of an adult pair is assumed to constitute a wintering ground unit. The regions are considered independent because of the fidelity of SP and because of the small dispersal range of the species ~\citep{colwell07,stenzel07,fwc2010,convertino10hur}. For PP and RK the three sites can be considered independent because these shorebirds use the areas as wintering and stopover areas for a limited period during which the inter-site movement was observed to be very limited. We considered all the renourishment projects that happened in every region in the previous year or before the nesting season in the same year, regardless of the size of the renourishment project. Specifically we considered all the areas that in the period 2002-2010 were subjected to at least one renourishment event. Those areas were considered as the potential breeding/wintering regions in the Bayesian inference. The occurrence of a nesting or a wintering ground is then checked in these regions. Let $Y$ be a random variable having a value of one if the region is not a nesting or wintering ground and zero otherwise and $X$ be a random variable having a value of one if the region was affected by a beach nourishment in the previous or in the same year and zero otherwise.  Then the odds ratio (OR) of a region not being a nesting or wintering ground given at least one renourishment event the year before or the same year relative to the region not being a nesting ground following a year without a renourishment is given by:

\begin{equation}
\hbox{OR} = \frac{P(Y=1|X=1) P(Y=0|X=0)}{P(Y=0|X=1) P(Y=1|X=0)} = \frac{\pi_1(1-\pi_0)}{\pi_0(1-\pi_1)}.\label{eq}
\end{equation}

The likelihood function given by the product of binomial distributions $Y_1 \sim \hbox{binomial}(\pi_1, n_1)$ and $Y_0 \sim \hbox{binomial}(\pi_0, n_0)$.  Assuming beta priors for the probabilities $\pi_0, \pi_1$ with parameters $(a_0, b_0)$ and $(a_1, b_1)$, our posteriors are given by $\pi_0|y \sim \hbox{beta}(y_0+a_0, n_0+b_0-y_0)$ and $\pi_1|y \sim \hbox{beta}(y_1+a_1, n_1+b_1-y_1)$. Assuming a uniform prior on the distributional parameters, we simulate posterior probabilities directly from the posterior distributions and compute the odds ratio as the fraction of the odds of an ``empty ground'' in the breeding season following a year with at least one beach nourishment project, to the odds of an ``empty ground'' in the season following a year without a beach nourishment. The Bayesian approach was performed for SP in the breeding and wintering season, and for PP and RK for the wintering season. 

 
 \section{Results and Discussion} \label{results}

As for the TER-s shorebirds analyzed, the Tyndall area hosted about 40 \%, 25 \%, and 22 \% of the SP, PP, and RK populations. Additionally for the Wilson's Plover (competitor of the SP) 83 \% of the population was in the Peninsula in the breeding season, confirming the importance of those Florida west coast Gulf beaches. The high presence of the WP and SP in those areas can explain the high favorability of the Panhandle for these shorebirds during the breeding season. Pensacola and Eglin areas hosted on average 8-10 \% of the SP population in the winter and in the summer season. For PP and RK those areas represent less than 1 \% of the population. The Peninsula and the Atlantic coasts are the main wintering grounds for the migratory PP and RK. The PP Peninsula population was 38 \%, and the Atlantic population was 33 \%. For the RK the proportions were 55 \% and 20 \% for the Peninsula and the Atlantic. Broad-scale estimates of the fledge-rate for Snowy Plovers nesting in Florida was estimated from data (Table \ref{table1}), however juvenile survival rates and adult survival remain still unknown. According to the trend of the average number of fledglings, an increase in the breeding population of SP is expected, supposedly because the increasing care in their conservation.\\

\begin{figure}
\begin{center} 
\vspace{-40pt}
\advance\leftskip-1.5cm
\includegraphics[width=16cm]{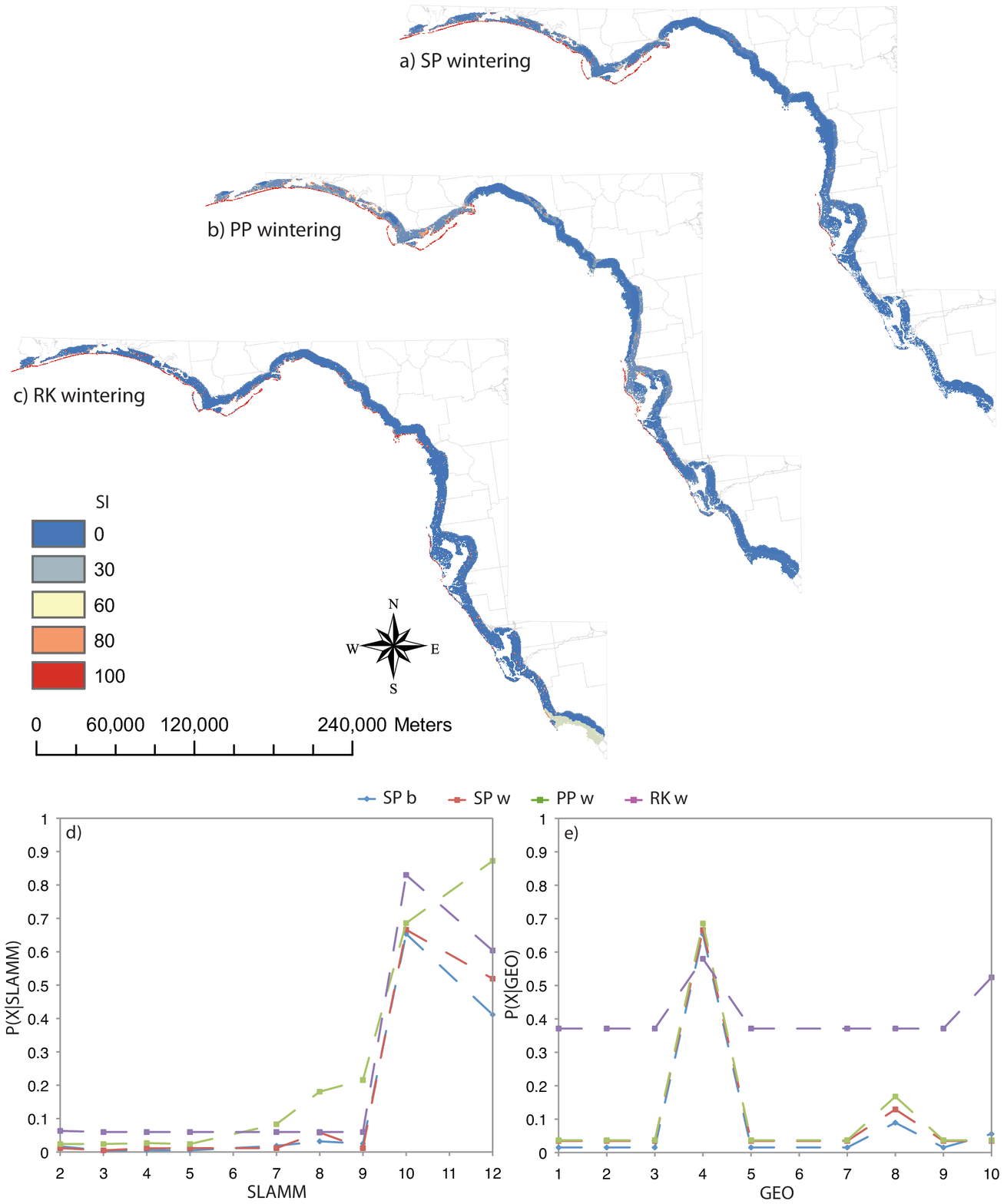}
\caption[h]{} \label{fig4}
\end{center}
\end{figure}

Figure \ref{fig4} reports the suitability index maps (SI) for the wintering season of SP (Fig. \ref{fig4}, a), PP (Fig. \ref{fig4}, b), and RK (Fig. \ref{fig4}, c) in the Panhandle-Big Bend-Peninsula region. We decided to model PP and RK in the same geographic domain where the range of the SP occurs in order to perform a comparative analysis for the habitat use of the studied shorebirds. The constraints of the habitat suitability model ({\sc MaxEnt}) are the adult-pairs occurrences in 2006, while the explanatory variables are the land-cover translated into SLAMM habitat classes \citep{chuagor10,convertino10scalesbirds}, and the geology GEO classes \citep{convertino10scalesbirds}, at a resolution of 120 m. The maximum entropy principle method calculated the probability map (from 0 to 1) assuming a regularization parameter equal to one, pseudoabsences placed as in \citep{convertino10scalesbirds}, and 25\% of the occurrences as training sample. The breeding habitat suitability map for SP is reported in \citep{convertino10scalesbirds}. From the habitat suitability map that can be considered the predicted fundamental species distribution, we attributed the class Suitability Index SI=60 to every pixel whose probability values are $> \; 0.2$ (threshold value) of the species distribution model. In this way considering only the pixels for SI $\geq$ 60 the geographic range for the shorebirds analyzed is successfully reproduced \citep{convertino10time}. All the pixels for SI $<$ 60 are categorized as unsuitable. SI=60 is considered the lowest score associated with consistent use and breeding/wintering, SI=80 is the score typically associated with successful breeding/persistent wintering, and SI=100 is for the best habitat with the highest survival and reproductive success or stable wintering \citep{majka07,convertino10time}. Figure \ref{fig4} (d, e) shows the response function or conditional probability of presence as a function of the two explanatory variables, the land cover and the geology. The logistic prediction changes as each environmental variable is varied, keeping all other environmental variables at their average sample value. In other words, the response curves show the marginal effect of changing exactly one variable, whereas the model may take advantage of sets of variables changing together. From the response curves it is possible to note the similar habitat preferences of SP in the breeding (dashed blue line) and wintering (dashed red line) seasons. In the winter the SP seem to use more the ocean beaches (class 12) than in the breeding season as documented by \citep{lamonte02}. The Piping Plover has a habitat preference similar to that of the Snowy Plover. Results of Figure \ref{fig4} (b) (green curve) show that the PP also occupies scrub/shrub transitional marsh and salt marsh areas with higher probability than the SP (class 7, 8, and 9 respectively) as confirmed by observations \cite{elias00,smith04ok}. Results also show that the PP seems to use more the ocean beach than SP does in the winter (higher P(X|SLAMM=12)). The RK in the winter seems to prefer more estuarine beaches (class 10). However, relatively high values of the probability of occurrence are observed also for the other SLAMM classes in comparison with SP and PP. The presence of medium/fine alkali sand and silt is less a requirement for the RK. RK is very adaptable to any substrate and in particular the suitability is high also for peaty-substrate habitats (class 12 GEO) as reported by \citep{niles10}. The minimization of the uncertainty of these results has been obtained in \citep{convertino10uncert}, for example considering the positioning error of recorded occurrences and the spatio-temporal gaps between occurrences and land-cover maps. The differences in the habitat use among shorebirds in the winter underline the importance of careful restoration planning policies that try to accommodate the needs of all the sensitive species. From the habitat suitability modeling, it appears that the resident SP is the most sensitive of the three species in relation to the habitat use (ocean and estuarine beaches) when subjected to variations due to renourshiment events. Piping Plovers have a very similar habitat use in the wintering season when they migrate to the coast of Florida; however they seem to be more resilient to habitat variations than Snowy Plovers because of their wider habitat preferences. Red Knots are the least sensitive species in relation to renourishment projects that modify the estuarine/ocean beaches, since they show the broadest spectrum of habitat preferences.\\

Figure \ref{fig5} reports the observed distribution of SP nests and SP, PP, and RK adult-pairs by year for the Pensacola/Eglin, Tyndall, and Peninsula study areas. The renourishments in the 2005-2006 are represented in Figure \ref{fig5}, however in the analysis the whole 2002-2010 period of data is considered (Table \ref{table1}). In each plot the breeding and wintering distributions of SP, PP, and RK are reported in the same year or in the year following the replenishment events. Only few nesting and wintering grounds occur in the locations where beach renourishment events occurred the previous year or the same year. In the data available there is not a complete information about the detailed timing of each intervention. As a result it is not possible to fully understand if the renourishments were performed during the wintering season or, less likely, during the breeding season and what was the duration of the project. However this information was partially compiled using the renourishment data reported in the Supplementary Material. The purpose of the current study was to infer the feedback between renourishment and TER-s shorebirds considering their spatial occurrences. Consequently we can not distinguish the direct or long-term effect of the renourishment on shorebird species. Renourishments events occured also during the beginning of the SP breeding season along the Florida Panhandle. Those interventions surely disturbed directly the site-selection processes. SP tend to have high-fidelity to nesting sites. Moreover, they are solitary nesters, unlike colonial waterbirds, and direct disturbances during the breeding season can seriously increase the fragmentation of the suitable habitat that affects their distribution and nesting success. The integrity and extension of a species' habitat are features that affect the survivability of the single individuals and the extinction risk of the whole population.   \\

Figure \ref{fig6} show the posterior probability of absence $P(A > a)$ of the odds ratio of not being a SP breeding ground, or a SP, PP, and RK wintering ground, based on the historical data. For SP over the years 2002--2010 we have one case (2004) in the Panhandle in which there was not any renourishment activity and there were more than two nesting grounds. In the same year (2004) for the Peninsula there were 2 renourishment projects and no nesting grounds. A value of $P(A > a)$ above one indicates a relatively higher probability of a region not being a nesting ground following a year with a renourishment.  The distribution is skewed to the right with a mode of about 1.7.  The median value of the odds ratio indicates that it is 2.5 times more likely that a region will not be a nesting ground for a SP following a year with a  replenishment event (dashed solid curve). A 90\% confidence interval for the odds ratio is (2, 30).  Specifically, for the SP we have $y_1$ = 12 cases of SP nesting grounds (as regions having at least 10 nest counts) over a sample of $n_1$ = 46 breeding seasons (counting separately all the renourished areas in 2002-2010) that were exposed to a beach nourishment the year before or the same year, while we have $y_0$ = 30 cases of nesting grounds over a sample of $n_0$ = 45 breeding seasons that were not exposed to a previous year/same year renourishment. Assuming a uniform prior ($a_0 = a_1 = b_0 = b_1 = 1$), we simulate 10$^4$ Monte Carlo posterior probabilities directly from the posterior distributions $f(\pi_0|y)$ and $f(\pi_1|y)$ and compute the OR as given in Eq. \ref{eq}.  Statistics on the OR are derived directly from the posterior samples. The results for the historical wintering distribution indicate that it was 2.3, 3.1, and 0.8 times more likely that a region was not a wintering ground following a year with a replenishment event for the SP (solid red curve), PP (solid brown curve) and RK (solid green curve) respectively.  The fit of the calculated probability is made by a lognormal distribution with different shape parameter. The posterior probability of absence for the breeding season is stable for values of the cutoff in considering an area a breeding ground of $10 \pm 4$ nests. There have been speculations that biological mechanisms induce log-normal distributions \citep{koch66}. In the majority of plant and animal communities, the abundance of species follows a truncated log-normal distribution \citep{sugihara80,limpert01}. Our conclusion is that regardless of the lack of detailed information about renourishment projects we are confident in assessing the negative feedback between past renourishment projects and SP, PP, and RK at the local-scale and at the macroscale for the whole region considered. The RK appears to be the least affected shorebird by beach renourishments. Our goal is to emphasize ecological sustainable restoration projects that take into account the habitat preferences of resident and migratory species. For example, the use of submerged geotexile groins that trap the sand nearshore, has no impact on the existing beach/ocean habitat communities and increases the beach area naturally. A successful application of geotexile groins has been tested in the Charlotte county shores in Florida also during the devastating 2004-2005 tropical cyclone season \citep{geotexile2005}.\\

The large-scale prediction of the habitat suitability for the whole SP Florida range \citep{chuagor10,convertino10scalesbirds,convertino10time} coupled with the local niche analysis as a function of physiognomic and biotic variables (here assumed to be invariant to climatic changes), environmental variables (e.g. tropical cyclones \citep{convertino10hur}), and human interventions (e.g. renourishments) will significatively help the estimation of the extinction and decline risk of Snowy Plovers in population viability models. These quantitive studies will enhance the adoption of effective conservation policies, e.g. through multi-criteria decision analysis, for the conservation of imperiled species.

\section{Conclusions} \label{concl}

The identification of habitat cues is critical for appropriate ecosystem management aimed to the conservation of rare and declining species. A comprehensive Hutchinsonian description of species is a necessary step in accomplishing this goal, combined with the understanding of the species's biogeographical distribution. Local scale analysis and macroscale studies need to be combined to adopt efficient conservation policies. 
Here we focused on the relationship between resident and migratory Threatened, Endangered and at-Risk (TER) shorebirds and renourishments along the Florida Gulf coast. In particular the focus was on the Snowy Plover, a resident state-designated threatened shorebird. In comparison we analyzed the Piping Plover, a migratory federally-designated threatened shorebird and the Red Knot which is a least-concern shorebird in Florida. The analysis was performed on nest occurrences and not on nesting success or chick survival data, which are not available. However the observed spatio-temporal correlation between renourishments and nest/adult-pairs of shorebirds can provide insights into the causes of the reduced number of nests after renourishment interventions. The following conclusions are worth mentioning.

\begin{itemize}

\item Based on the 2006 wintering and breeding counts Tyndall is confirmed to be the hotspot of the shorebird species richness in Florida both for resident and migratory birds.  The differences in the breeding and wintering distribution of the SP population (in the Panhandle 80 and 60 \% in the breeding season and in the wintering season; in the Peninsula 20 and 40 \% respectively) and the almost unaltered total pairs count can potentially confirm the predicted movement of SP ($\sim 20 \%$) to the lower and warmer latitudes of the Peninsula beaches during the winter. However SP from other Gulf state beaches have been sporadically observed in Florida. The result confirm the observations for the Snowy Plover subpopulation in the West coast of the USA \citep{stenzel94}, in which the inter-seasonal dispersal happens for longer distances than within seasons. Renourishment interventions need to carefully preserve Tyndall as focal area for shorebirds richness. Moreover dispersal patterns have to be considered in order to reduce the potential direct disturbance of renourishments. Nonetheless further studies about the shorebirds movements patterns in Florida are needed;

\item During both nesting and brood-rearing stages of breeding, Snowy Plover selection of habitat and productivity are influenced by a combination of abiotic and biotic factors including human disturbance, predator abundance, prey availability and the physical features of the habitat. However, we did not study the factors influencing different breeding stages due to the lack of data. The habitat suitability model based on the principle of maximum entropy ({\sc MaxEnt}) \citep{phillips06,phillips08b}) did not observe consistent differences between the habitat preference of SP in the breeding and in the wintering season. In the winter the SP seems to utilize more the ocean beaches than in the summer, possibly because the brood-rearing habitat is less used than during the breeding. In the wintering season the habitat use of the migratory PP is very similar to that of the SP. In contrast, the RK seems to have a larger spectrum of habitat preferences. The PP utilizes the ocean beaches more than the SP. However, it also prefers scrub/shrub transitional marsh and salt marsh areas. The RK is observed occasionally also on peaty-substrate banks. We did not consider any possible direct interaction (such as interspecies density-dependence) between the shorebird species. PP and RK seem more resilient to the effects of renourishment projects (that modify estuarine and ocean beaches) that are not focused on the conservation of the wildlife habitat because their habitat preference is wider than SP;

\item We argue that the decrease in the documented nest abundance of SP resulted from an altered cross-sectional beach profile which is not favorable for nesting and foraging. As a consequence the nesting success is reduced because of habitat modifications and possible direct disturbance of renourishment interventions. The profile subsequently improved in later seasons as the beach equilibrated to a more natural slope and surface roughness more significantly. A negative feedback was found between SP, PP, and RK and historical renourishment projects. Results based on a spatio-temporal Bayesian inference show that it was 2.1, 3, and 1 times more likely that a region was not a wintering ground following a year with a replenishment event for the SP, PP and RK, respectively. Despite the fact that the inference for the TER shorebirds in the winter is based only on the 2002 and 2006 census we believe it is significant and in agreement with the observed behavior. Considering  the census of nine breeding seasons and the renourishment events in the same period, the median value of the odds ratio indicates that it was 2.5 times more likely that a region was not a nesting ground for a SP in the same or in the season following a renourishment event. The higher median of the OR in the breeding season indicates that the renourished beaches were more altered in the summer season immediately after the renourishments (performed mostly in the winter). The physiochemical properties of the dredged material for the renourishment were equivalent to those of the existing in-situ sand \citep{lott09,finalrep2009,nourish10}. It can therefore be concluded that the ecogeomorphological alteration of the habitat was the only cause of the diminished occurrence of shorebirds in the replenished sites. Natural processes such as overwash and tropical cyclones shaped the renourished beaches, bringing them to their non-altered configuration, for e.g. with the creation of ephemeral pools and bay tidal flats that constitute the favorite habitat for SP and PP \citep{convertino10hur};

\item We emphasize the importance of an a-priori planned ``ecological sustainable renourishment'' (e.g., performed  by submerged geotexile groins that preserve the beach cross-section profile) that consider the triality among dredging costs, protection of the coastal structures, and the potential income from the value of the preserved biodiversity and the enhanced recreational activities on the extended beaches. Renourishment projects designed to create high quality brood-rearing habitat characterized by sparse low vegetation, and the maintaining of high-prey foraging habitat is an important part of shorebirds conservation.

\end{itemize}

\vspace{1cm}

\textbf{Acknowledgments} 
The authors acknowledge the funding from  the Strategic Environmental Research and Development Program (SERDP) for projects SI-1699 and SI-1700. The authors also gratefully acknowledge the work of the Florida Fish and Wildlife Conservation Commission and the Florida Shorebird Alliance (employees, researchers, volunteers, and land owners) for collecting the SP data and making them available to the public. Chris Burney and Patricia Kelly (FWC) are acknowledged in particular. An anonymous reviewer is greatly acknowledged for his/her comments that significantly improved the manuscript. J.B. Elsner (Florida State University) is gratefully acknowledged for the insights regarding the model. Figure S2 (a) is from a field survey of R.A. Fisher, and Figures S2 (b, c) from a field survey of M. Convertino. The authors declare to have used the data only from the cited sources. The extensive assistance of  the Eglin AFB  personnel is particularly acknowledged. There are not conflicts of interests to declare. Permission was granted by the USACE Chief of Engineers to publish this material. The views and opinions expressed in this paper are those of the individual authors and not those of the US Army, or other sponsor organizations.





\clearpage

\thispagestyle{empty}

\begin{center}
{\bf Table Captions}
\end{center}

\begin{figure}
\begin{center} 
\vspace{-40pt}
\advance\leftskip-1.5cm
\includegraphics[width=14cm]{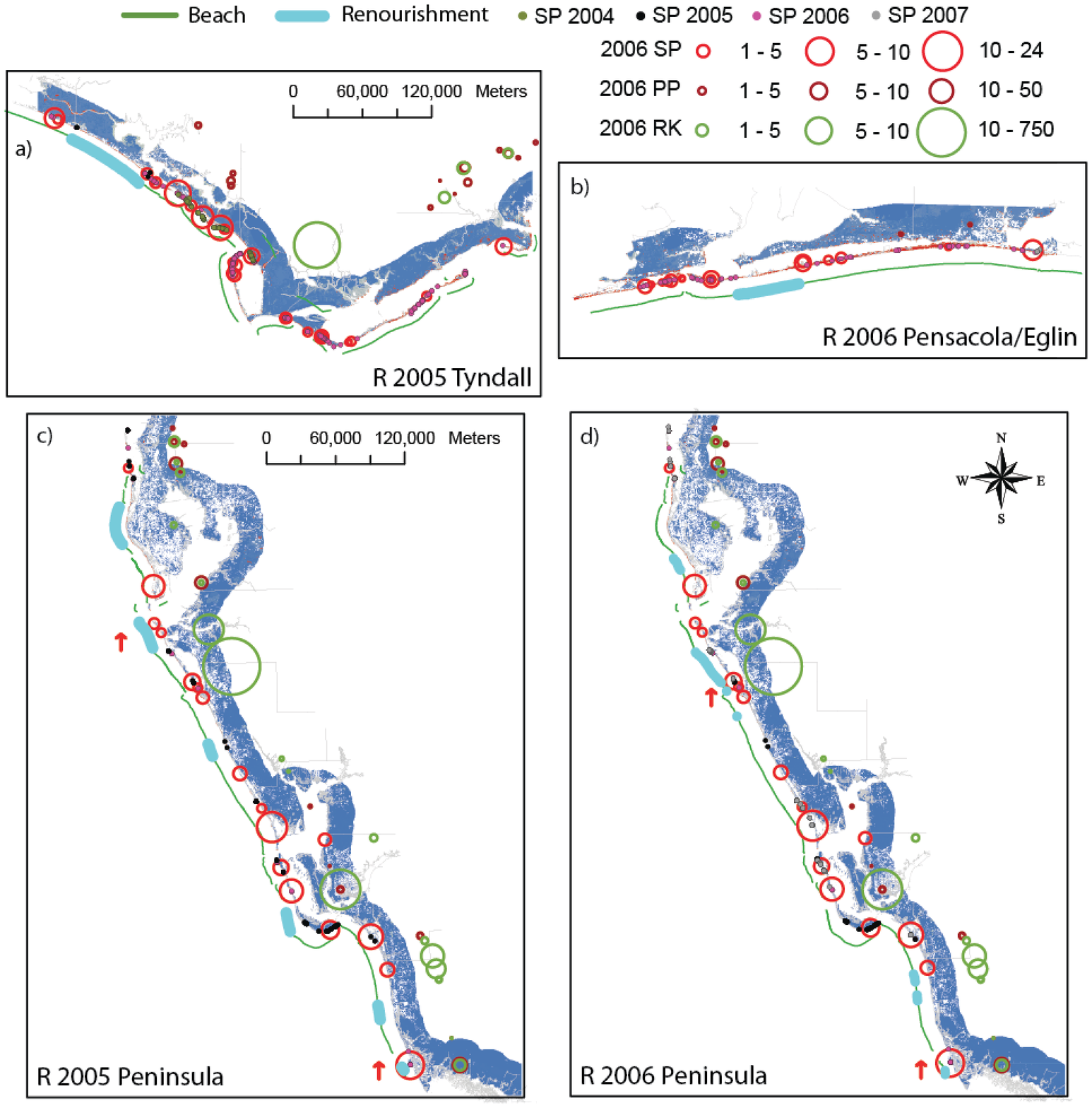}
\caption[h]{} \label{fig5}
\end{center}
\end{figure}
\begin{figure}
\begin{center} 
\vspace{-40pt}
\advance\leftskip-1.5cm
\includegraphics[width=10cm]{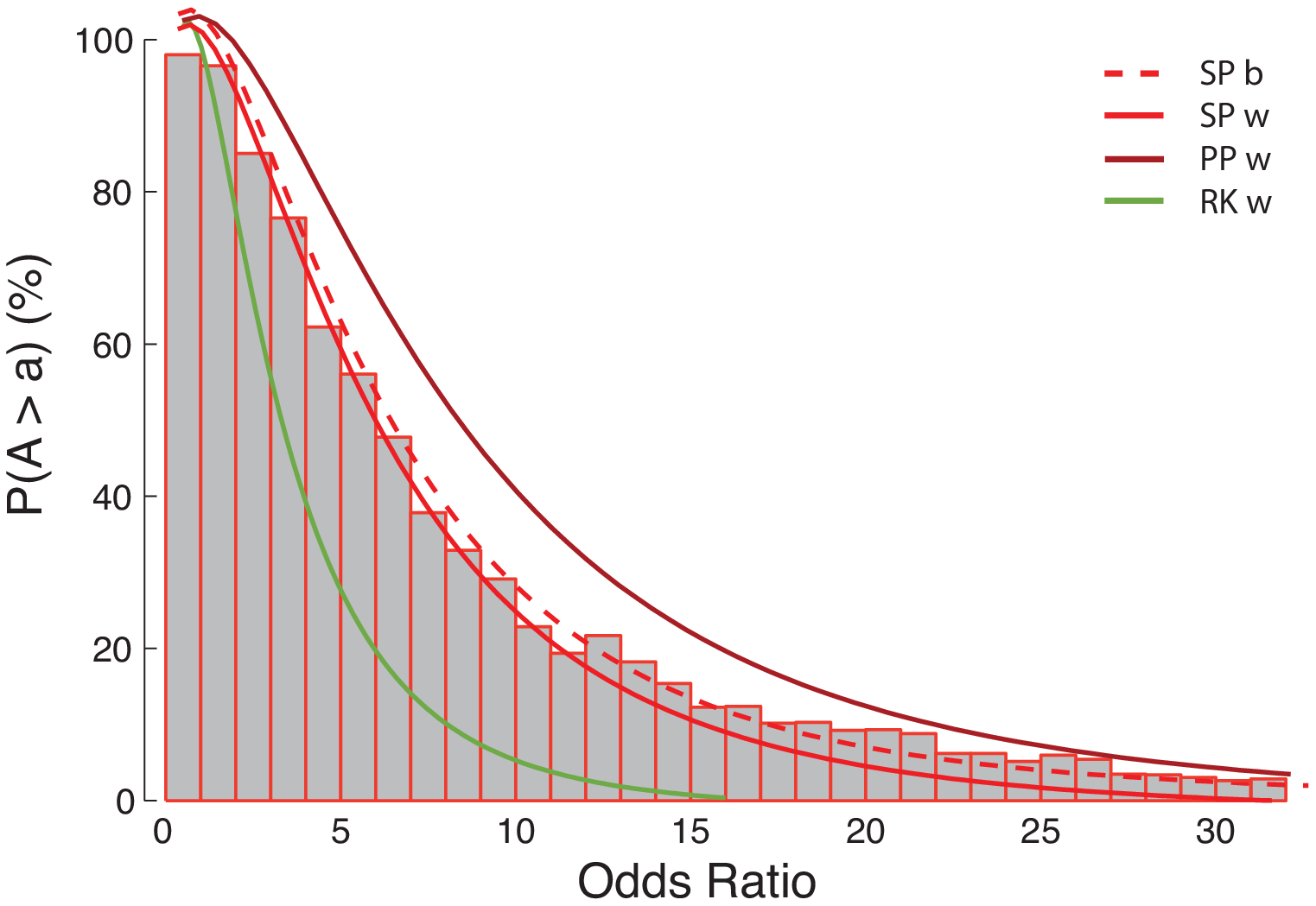}
\caption[h]{} \label{fig6}
\end{center}
\end{figure}

{\bf Table 1.} Breeding ($b$) ~\citep{fwc2010,fsa2010} and wintering ($w$) \citep{ippc09} counts (SP nest and adults pairs respectively) for the Panhandle, Peninsula and the whole populations in the seasons 2002-2010. A SP pair is considered to exist for every nest count. The number of renourishment events \citep{fsu10}, $r$ (P and T stand for Pensacola/Eglin, and Tyndall), refers to the renourishments made the same year or the year previous to the breeding and wintering season. We considered as potential breeding/wintering regions in the Bayesian inference all the areas that in the period 2002-2010 were subjected to at least one renourishment event. The number of fledglings is estimated from the observed counts. The data in the gray background are those considered for the Bayesian inference in the seasons 2005-2006 (Figure \ref{fig5}).  \\


\begin{center}
{\bf Figure Captions}
\end{center}

{\bf Figure 1.} (a) Map of the Panhandle-Big Bend-Peninsula (PBBP) study area along the Gulf coast of Florida, and closeups of the three focal study sites (Pensacola, Eglin, and Tyndall (Apalachee Bay)) in which there is a high density of military areas, federal reserves and state parks. The red dots refer to the 2006 Snowy Plover (SP) breeding census performed along the Florida coastline ~\citep{fwc2010,fsa2010}. The gold dots are the observed SP adult-pairs in the winter season \citep{ippc09}. The blue dots are the nests of the Wilson's Plovers ~\citep{fwc2010,fsa2010} (see Supplementary Material). The critical beaches are depicted in red based on the 2009 Critical Erosion Report \citep{criterod10}. (a), (b), and (c) are the 2010 satellite images of the selected study sites with the US military bases involved in the study delineated in red. (b) wintering distribution of Piping Plover (PP) and Red Knot (RK) in 2006 \citep{ippc09}.   \\

{\bf Figure 2.} Suitability Index maps derived from the average over 30 habitat suitability realizations calculated by {\sc MaxEnt} for the Snowy Plover (SP) (a), Piping Plover (PP), and Red Knot (RK) as a function of the land-cover, the geology layers \citep{convertino10scalesbirds}, and the 2006 winter adult-pairs occurrences. The conditional probability $P(X|Y)$ to find a nest or an adult-pair is plotted as a function of the continuous explanatory variable $Y$ at resolution 120 m, that is the land-cover translated into SLAMM habitat classes \citep{chuagor10,convertino10scalesbirds} (d), and the geology GEO (e), for the model run keeping all other environmental variables at their average sample value. $X$ is a SP nest in the SP breeding region or a SP, PP, and RK adult-pair in the wintering season.  \\

{\bf Figure 3.}  Distribution of SP nest sites (dots) and SP, PP, RK adult-pairs sites (circles proportional to the adult-pairs abundance) by year for Tyndall (a), Pensacola/Eglin (b), and Peninsula areas (c, d). The renourishments in the 2005-2006 period are represented. In each plot is reported the breeding and wintering distribution of SP, PP, and RK in the same year or in the year following the replenishment events. In the background the habitat suitability is represented for the buffer of 10 km from the coastline \citep{convertino10scalesbirds}. A dot may represent more than a single nesting site. Within the plot the text indicates the renourishment $R$ in the previous year that is represented by a continuous light-blue line in the map. Red arrows indicate the sites with positive feedback SP-renourishment. \\

{\bf Figure 4.} Posterior probabilities of absence $P(A > a)$ of the odds ratio for SP in the breeding season, and SP, PP, and RK in the wintering season.  The odds ratio is the ratio of the odds of a nesting or wintering ground in the spring following a year with at least one renourishment event to the odds of a nesting or wintering ground in the spring following a year without a renourishment intervention.  For the breeding SP the median odds ratio is 2.5 and the mean is 4.9. For the wintering SP, PP, and the RK the median odds ratio is 2.3, 3.1 and 0.8 respectively. The maximum likelihood estimate is a lognormal distribution with different values of the shape parameter (histogram only for SP) and the coefficient of determination for SP, PP, and RK is on average R$^2=0.92$.

\newpage

\clearpage
\begin{table}
\caption{} \centering \label{table1}
\vspace{0.2 cm}
\begin{tabular}{l|ccc|ccc|cc|c}
\hline \hline

 &\multicolumn{8}{c|}{Pairs}  \\

\multicolumn{1}{c|}{Year} & \multicolumn{3}{c|}{Panhandle} & \multicolumn{3}{c|}{Peninsula} & \multicolumn{2}{c|}{Whole} & \multicolumn{1}{c}{Fledglings} \\
 &  $b$ & $w$ & $r$ & $b$ & $w$ & $r$ & $b$ & $w$   \\
\hline

$2002$ & 128 & 228 & 1 (P)  & 65 & 103 & 1 & 193 & 332 & -  \\

$2003$ & 9 & - & 1 (P) & 1 & - & 4 & 10 & - & -  \\

\rowcolor[rgb]{0.9,0.9,0.9} $2004$ & 57 & - & - & - & - & 2 & 57 & - & - \\

\rowcolor[rgb]{0.9,0.9,0.9} $2005$ & 4 & - & 2 (P, T) & 93 & - & 6 & 97 & - & 1.082 \\

\rowcolor[rgb]{0.9,0.9,0.9} $2006$ & 235 & 175 & 2 (P, T) & 68 & 137 & 7 & 303 & 312 & 1.075 \\

 $2007$ & 48 & - & 1 (P) & 96 & - & 6 & 144 & - & 1.524 \\

 $2008$ & 394 & - & 1 (P) & 78 & - & 1 & 472 & - & 2.000 \\

$2009$ & 1051 & - & 2 (P) & 195 & - & 1 & 1246 & - & 0.860 \\

$2010$\footnotemark[1] & 241 & - & 2 (P, T) & 66 & - & 5 & 307 & - & 1.390 \\

\hline \hline
\end{tabular}
\begin{tablenotes}
\item[1] $^1$Data updated to the 30 July 2010 of the breeding season.
\end{tablenotes}
\end{table}


\newpage
\clearpage \thispagestyle{empty}

%
%
%

\newpage
\clearpage \thispagestyle{empty}

\newpage
\clearpage \thispagestyle{empty}

\newpage
\clearpage \thispagestyle{empty}

\end{document}